\documentclass[apj]{emulateapj}
\usepackage{amssymb,amsmath}
\usepackage{color}

\definecolor{darkgreen}{rgb}{0.0,0.6,0.0}

\newcommand{\be}{\begin{equation}}
\newcommand{\ee}{\end{equation}}
\newcommand{\bea}{\begin{eqnarray}}
\newcommand{\eea}{\end{eqnarray}}
\def\se#1{Section~\ref{sec:#1}}

\def\Fig#1{Figure~\ref{#1}}
\def\Table#1{Table~\ref{#1}}

\def\kep{{\em Kepler}}

\begin{document}

\title{Measuring the Number of M-Dwarfs per M-Dwarf \\ Using \kep\ Eclipsing Binaries}

\author{Yutong Shan$^{1}$, John A. Johnson$^{1}$, \& Timothy D. Morton$^{2}$}
\altaffiltext{1}{Department of Astronomy, Harvard University, 60 Garden Street, Cambridge, MA 02138}
\altaffiltext{2}{Department of Astrophysical Sciences, Princeton University, 4 Ivy Lane, Peyton Hall, Princeton, NJ 08544}

\begin{abstract}
We measure the binarity of detached M-dwarfs in the \kep\ field with orbital periods in the range of 1--90 days. \kep's photometric precision and nearly continuous monitoring of stellar targets over time baselines ranging from 3 months to 4 years make its detection efficiency for eclipsing binaries nearly complete over this period range and for all radius ratios. Our investigation employs a statistical framework akin to that used for inferring planetary occurrence rates from planetary transits. The obvious simplification is that eclipsing binaries have a vastly improved detection efficiency that is limited chiefly by their geometric probabilities to eclipse. For the M-dwarf sample observed by the \kep\ Mission, the fractional incidence of eclipsing binaries implies that there are $0.11 ^{+0.02} _{-0.04}$ close stellar companions per apparently single M-dwarf. 
Our measured binarity is higher than previous inferences of the occurrence rate of close binaries via radial velocity techniques, at roughly the 2$\sigma$ level.  
This study represents the first use of eclipsing binary detections from a high quality transiting planet mission to infer binary statistics. Application of this statistical framework to the eclipsing binaries discovered by future transit surveys will establish better constraints on short-period M$+$M binary rate, as well as binarity measurements for stars of other spectral types.

~

\end{abstract}


\section{Introduction}\label{sec:intro}

Our existence in a planetary system around an isolated star makes it easy to forget that single, G-type stellar systems are not the norm throughout the Galaxy. For example, among the FGK dwarfs in the Solar neighborhood, only 54\% are single \citep{rag10}. Furthermore, studies of the stellar populations of open clusters \citep[e.g.][]{cha03} and the Solar neighborhood \citep{hen06, win15} have demonstrated that over 70\% of the hydrogen-burning stars in the Galaxy are in fact M-dwarfs, i.e. stars whose masses are below $0.6 M_\odot$. Multiple-star systems and low-mass, M-type stars are therefore common outcomes of star formation.

Observed metrics of multiplicity across various stellar types provide important constraints for star formation and dynamical evolution theories (see review by \citealt{dk13} and references therein). For the burgeoning field of exoplanets, however, there are several other reasons to study the binarity of M-dwarfs.

For one, it turns out binary stars can host planetary systems. $\alpha$ Cen B is a particularly noteworthy example, with an Earth-like planet candidate in a short-period orbit around a member of a triple--star system \citep{dum12, pla15}. In fact, it may be quite common for planets to orbit one component of a widely-spaced binary system. For instance, \citet{ngo15} find that, for stars hosting hot Jupiters, the incidence of stellar companions is $\sim50$\%. Planets have also been discovered in circumbinary configurations. At first considered only exotic possibilities, circumbinary planets are a new class of exoplanetary system discovered by the \kep\ Mission, with one or more planets orbiting exterior to a pair of stars \citep{doy11, wel12, oro12, kos14, wel14}. 

M-dwarfs are also prolific planet hosts, particularly for planets with radii and masses less than those of Neptune. \citet{ms14} find that there are on average $2.00 \pm 0.45$ planets per M-dwarf with periods less than 150 days, based on the statistics of \kep\ transit detections, with a peak in the radius distribution near $\sim1 R_\oplus$. \citet{dc15} find that, conservatively, there exist 0.18 Earth-sized planets per M-dwarf habitable zone. Owing to their favourable brightness contrast and the close-in location of their habitable zones, planets around M-dwarfs offer great potential for discovery and characterization with current and future missions, such as MEarth \citep{mearth}, K2 \citep{howell14}, TESS \citep{ric14}, and upcoming RV surveys \citep[cf also][]{dc13}.   

Binaries can also confound the effort to search for planets. An eclipsing binary system in the background of a target star could mimic foreground planet transit signals, forming an important source of astrophysical false positives for transiting planet surveys \citep{tor11,mj11}.

Given these considerations, M-dwarf multiples constitute an important subclass of objects for stellar and exoplanetary science. 
The first large multiplicity study targeting low-mass primaries was conducted by \citet{fm92} (hereafter FM92). In this seminal investigation, the authors combine surveys based on a variety of techniques, including radial velocities, high-contrast imaging, and interferometry, each sensitive to a particular orbital separation regime. In their calculation of multiplicity statistics of systems with M-dwarf primaries\footnote{Hence the secondary must either be an M-dwarf or substellar.}, FM92 measure frequency distributions in semimajor axis and mass ratio. Overall, they find that $42\% \pm 9\%$ of the M-dwarfs studied are in multiple systems within the separation range of 0.04 to $10^4$ AU. 

Since FM92, the occurrence statistics of low-mass stellar systems have been the subject of various works. Some highlights are: 
orbital elements of spectroscopic binaries among local early- to mid-M-dwarfs \citep{may00,udr00},
volume-limited study of M-dwarfs within 9pc \citep{del04}, late-M-dwarf primaries \citep{all07}, visual M-dwarf systems \citep{ber10, jan12, jod13, wd15}, young M-dwarf primaries \citep{bow15}, X-ray bright M-dwarfs and spectroscopic binaries \citep{shk10}, and very short-period M-dwarf binaries \citep{cbk12}.  
These studies rely on either direct imaging or spectroscopic/radial velocity (RV) techniques. The former has limited sensitivity to low projected component separations and contrasts, whereas the latter theoretically makes up for completeness for very closely orbiting systems, but requires bright targets. 
The overall multiplcity for M-dwarf primaries found by these studies lie between 20\% and 40\%.   

The closest separation bin considered in FM92 is 0.04~AU to 0.4~AU. For the typical M-dwarf masses involved, such a range of semimajor axis corresponds to orbital periods of approximately $P \lesssim 90$~days. This very tight orbital range of M-dwarf binaries is interesting yet enigmatic territory. 
It is interesting in providing constraints to theories of tight binary formation (e.g. \citealt{bb94b, bbb02}). It is also a discovery space for transiting circumbinary planets, for which the binary orbital period needs to lie well within the orbital period of any detectable planet. Currently, the sensitivity for the latter is no more than a couple of years, limited by survey duration. 
Furthermore, background EBs with periods comparable to the time baseline of high-precision transiting planet surveys are the most likely astrophysical false positives. Hence, close binary occurrence rates form a crucial input into calculations of planet statistics \citep{fre13}.

However, M$+$M short-period binaries have been enigmatic because they are relatively difficult to identify visually due to faintness and spatial resolution limits. Furthermore, predictions for the abundance of such systems are inaccessible by state-of-the-art hydrodynamic simulations of star formation, also due to resolution limits subjected to computation budget \citep{bate09, bate12}. To date, all empirical constraints on the statistics of these systems derive from RV analyses \citep{fm92,udr00,del04,shk10,cbk12}. 

FM92 find precisely 1 M-dwarf companion within 0.4 AU among the 62 M-dwarfs subjected to their search. Correcting for completeness, they conclude that, in this range of orbital distance, the occurence rate of M-dwarf binaries is $1.8\% \pm 1.8\%$. 

Recent work by \citet{cbk12} (hereafter CBK12) provides an alternative measurement. They examine spectra from Sloan Digital Sky Survey (SDSS) M-dwarfs to look for RV variations that reveal the presence of close stellar companions, thereby deriving an estimate on the binarity for orbits with $a <0.4$~AU. 
They report the detection of 22 binary candidates among the 1452 M-dwarf targets examined. Correcting for detectability, CBK12 infers a close binary fraction of 3-4\% among the M-dwarfs.



The NASA \kep\ Mission has been highly successful at advancing exoplanet science \citep{bor10}. Throughout its primary mission baseline of 4 years, it collected continuous and precise broadband photometry from some 160,000 stars in a 100-square-degree-patch of the sky near Cygnus (field centre: RA = 19h22m40s, Dec = $+44^\circ 30' 00''$). The resultant light curve database has enabled the discovery of thousands of transiting planets. Furthermore, the variability in these light curves encode information on the activity, rotation, pulsation, and other characteristics of the stars themselves \citep[e.g.][]{bas13,hub14,mos14,jj14,ang15}, extending their utility beyond those for whom planetary systems are fortuitously inclined. Thus, while \kep\ may be conceived chiefly as an exoplanet mission, it has hailed a veritable golden age of stellar astrophysics. 

Eclipsing binary (EB) systems are a class of objects observed by \kep\ that lie at the intersection of transits and stellar astrophysics. Furthermore, the occurrence of EBs in a sample of targets directly reflects the underlying binarity rate, discounted by the geometric eclipse probability---a function chiefly of semimajor axis, stellar radii, and orbital eccentricity. The detection of EBs is a simplified version of exoplanet hunting---a stellar eclipse, even grazing, is generally a much larger signal than that of a transiting planet, and thus far less likely to be missed. In addition to its impressive list of planet detections, \kep\ has uncovered thousands of eclipsing binaries \citep{cou11, prs11, sla11, mat12}. Given the importance of double-lined spectroscopic EBs as the only model-free way of simultaneously measuring the precise radius and mass of stars, this is a valuable database in and of itself for testing stellar models, especially in the regime of low-mass stars where theory and observation are currently discrepant \citep[e.g.][]{lm08, tag10}.

Owing to the mission mandate of \kep, which aims to quantify the prevalence of Earth-like planets, the methodology for computing exoplanet occurrence statistics has become highly developed (e.g. \citealt{you11,how12,dc13,pet13,fre13,ms14,fm14}). Since stellar systems are akin to planet systems, the same framework can be transplanted to calculate stellar companion statistics, with several simplifications. The chief source of incompleteness ought to be due to non-transiting geometries. Since we can quantify the transit probability of a binary system given its scaled semimajor axis ($a/(R_1 + R_2)$), which is measured by \kep, each EB detection can be assigned a statistical weight to reveal the underlying population that is not in an eclipsing configuration. 


In this work, we explore what \kep\ can reveal about close (but detached) M$+$M-dwarf binarity statistics. We begin by defining our sample and describing the raw \kep\ data and how we conditioned them to our purpose (\se{datmet}). The scheme for culling the EBs is described in \se{cull}. The EB detections among the sample are listed in \se{validation}, and comparisons are made with previous studies in \se{comparelist}. Then we elaborate on the statistical method alluded to in the previous paragraph (\se{EBstats}). The overall {\emph{NSPS} -- Number of Stars per Star --} and its orbital period distribution are presented in \se{discuss}, followed by a discussion of the caveats, biases, and comparisons to previous studies. \se{conclusion} concludes with a summary of the findings.


\section{Sample \& Data}\label{sec:datmet}

\subsection{Sample} \label{sec:sample}

\begin{figure}[htb]
\center
\vspace{-10pt}
\hspace{-20pt}
\includegraphics[scale=0.6]{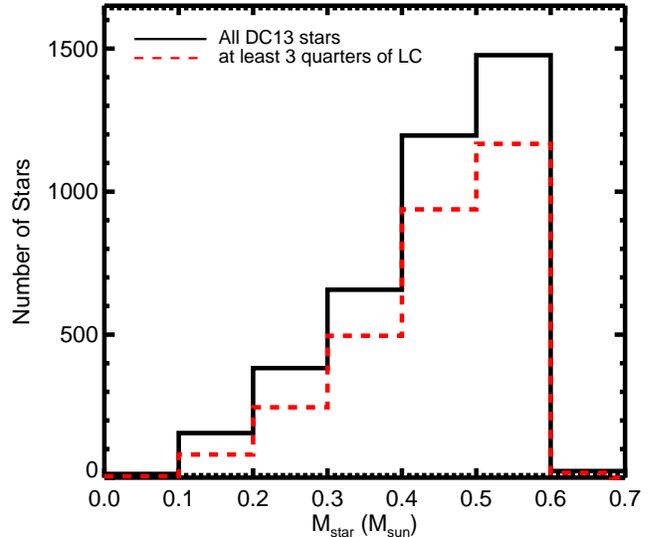}
\vspace{-10pt}
\caption{The stellar mass distribution of DC13 targets. The black solid line depicts all the stars compiled by the DC13 catalogue, of which there are 3905. This means they were observed for at least 1 quarter by \kep. Those stars with at least 3 quarters of LC observation, which is a requirement for stars used in our statistical study, are shown in red dashes. There are 2975 of them. Not surprisingly, the overall sample comprises predominantly early M-dwarfs.}
\label{dc-mstar}
\end{figure}

Our sample comprises the set of M-dwarf targets observed by \kep\ as identified by \citet[][henceforth DC13]{dc13}. DC13 also derive updated stellar parameters, such as masses and radii, estimated by comparing colours of model stars from Dartmouth stellar evolutionary tracks with that of the observed stars. For cool stars, these parameters should represent a vast improvement upon the original \kep\ Input Catalog (KIC, \citealt{bro11}). DC13 provides a total sample of 3905 early to mid M-dwarfs with $T_{\rm eff} \leqslant 4000$K, each observed for at least one \kep\ quarter. Distributions of stellar masses are shown in Figure \ref{dc-mstar}, exhibiting a heavy but expected Malmquist bias toward earlier spectral types, since the luminosity-mass relation is very steep ($L \sim M^5$) in this region of the H-R diagram.  

\begin{figure}[htb]
\center
\vspace{-15pt}
\hspace{-20pt}
\includegraphics[scale=0.61]{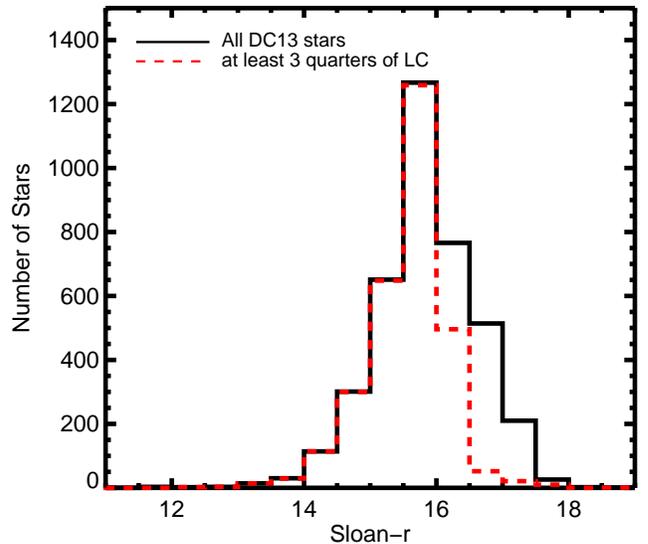}
\vspace{-10pt}
\caption{The Sloan-r magnitude distribution of DC13 targets. Colour scheme follows that in Figure \ref{dc-mstar}. The sample is neither volume- nor magnitude-limited.} 
\label{dc-sloanr}
\end{figure}

Note that the \kep\ sample is neither volume-limited nor, strictly speaking, magnitude-limited (see Figure \ref{dc-sloanr} for the Sloan-r magnitude distribution of the M-dwarf sample). As a result, the target selection is poorly defined since M-dwarfs were not the primary targets of the \kep\ Mission \citep{bat10a}. Nevertheless, we argue that this sample is {\emph{not unrepresentative}} of M-dwarfs in the Solar Neighborhood, particularly for early-M spectral types. 

The stars in the \kep\ M-dwarf sample did not have multi-epoch spectroscopic data at the time of selection, and as a result spectroscopic binaries were not intentionally excluded. Further, while distances to some stars in the \kep\ sample extend out to beyond a kpc, the M-dwarfs predominantly reside within $\sim200$~pc. Thus, we expect the occurrence of stellar companions measured by the rate of EBs in \kep\ to be representative of the binarity for M-dwarfs in the Solar Neighborhood (i.e. for relatively nearby stars within $\lesssim 20$~pc of the Sun).

\subsection{Data and Its Processing} \label{sec:data}
The majority of \kep's targets were observed in long-cadence (LC) mode, where each data point forming the light curve is an integration over $\approx29.4$~minutes. We downloaded the Pre-search Data Conditioning Simple Aperture Photometry (PDCSAP) \citep{stu12,smi12} for each target in the DC13 sample. This is a publicly available data product supplied by the \kep\ Team and made accessible via the Mikulski Archive for Space Telescopes (MAST \footnote{https://archive.stsci.edu/kepler/}). In addition to the standard bias and dark subtractions, field flattening, and cosmic ray treatment, PDCSAP strives to eliminate the systematic and instrumental effects with ``cotrending vectors,'' while preserving variations of astrophysical origin, such as rotational modulation of spots. 

To prepare the light curves for eclipse searches we perform a few additional tasks: we quarter-stitch following median-normalization and ``flatten'' the data over an appropriate timescale using piecewise polynomials. Over a chosen window of a set time interval, we perform a low-order polynomial fit to the light curve within this window, less any points that are $>5\sigma$ away from this functional fit, $\sigma$ being the standard deviation of the entire normalized light curve. This is iterated 5 times to reach a final polynomial fit for that window. Then the window slides over to the next segment and the process is repeated, though continuity on the shared boundary of the windows is not demanded. The resultant piecewise polynomial is subtracted from the original light curve. For the chosen polynomial order of 2 and window size of 1 day, we filter out smoothly varying features spanning longer than this timescale, representing a high-pass filtering process. This step rids typical periods of rotational modulation (fast-rotators excepted), while preserving any reasonable detached eclipse event lasting at most a few hours for the orbital period range of interest. For an early M$+$M EB on a 90-day circular orbit, the transit duration is $\sim$8 hours, and less for shorter orbital periods. 



\section{Methods} \label{sec:methods}

Of the 3905 M-dwarfs present in the DC13 sample, 2975 of them were monitored for at least 3 quarters by \kep, making them uniformly sensitive to sub-90-day light curve periodicities. With this sample of 2975 processed M-dwarf light curves, we can identify the eclipsing binaries among them.  The eclipse depth of a stellar binary is typically much larger than that of a planet transit. Of course, in calculating the expected eclipse depth for a binary system, one must not only be concerned with the radius ratio between the bodies, as one would in a planetary eclipse problem, but also the self-luminous nature of the eclipsing body. Hence, the radius and temperature of both components matter. 

We generate distributions of the expected primary eclipse depths for a realistic population of M+M EBs with DC13 M-dwarfs as primaries, accounting for grazing scenarios. To each primary we assign secondary masses according to an \"{O}pik's mass-ratio power law $f(q)\sim q^\gamma$, where $q\equiv M_s/M_p$, the ratio of the secondary to primary masses. Then, stellar masses are converted into radii and temperature using observationally calibrated relations from \citet{boy12}. Orbits are drawn from orbital element distributions as described in \se{EBstats}. For M-dwarf binaries below 5 AU, $\gamma$ may be as high as 2.7, but is highly uncertain \citep{dk13}. We adopt a more modest mass distribution power of 1 since it would result in a conservative eclipse depth distribution \footnote{A higher $\gamma$ corresponds to greater occurrence of higher-mass secondary companions, hence would give rise to larger eclipse depths overall}, and note that the outcomes presented below are robust to any reasonable variation in $\gamma$.   

\begin{figure}[htb]
\center
\vspace{-10pt}
\hspace{-1cm}
\includegraphics[scale=0.5]{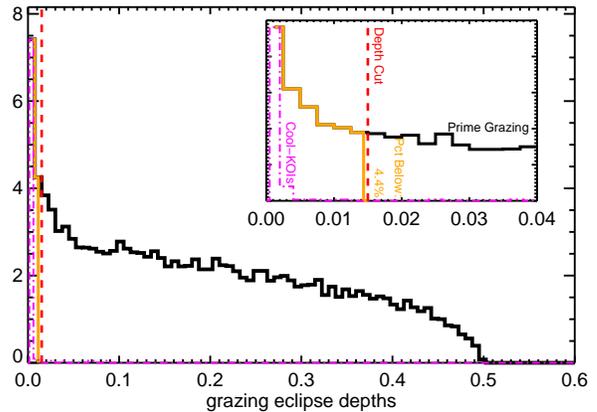}
\caption{A distribution of primary eclipse depths (i.e. fractional light blocked in eclipse) for a simulated population of M-dwarf EBs. Vertical axis is given in arbitrary units. The red dashed vertical line at 0.015 is the cut in depth we make to generate our first round of candidates. The magenta dash-dot histogram shows cool-KOI transit depths \citep{swi15}, approximated by $(R_p/R_\star)^2$. Less than 5\% of the EB cases lie below the selection criterion, contrasted with nearly all of the transiting planets around M-dwarfs.  
}
\label{mm-ed} 
\end{figure}

The distribution of eclipse depths is shown in Figure \ref{mm-ed}\footnote{In all the computations, limb darkening and effects such as beaming and ellipsoidal variations have been ignored \citep{fai-maz11}. Also, we do not consider substellar companions since they are intrinsically rare around stars generally, and in our sample specifically. See \se{completeness estimates}}. Based on our simulations, the eclipse depth can be up to 50\% for equal-mass constituents, and in $\>95$\% of the cases the depths are greater than 1.5\%, even in grazing geometries. Depths of this magnitude are easily identified in the light curves of even the faintest stars in our sample\footnote{Typical S/N for a 1.5\% single-point transit for M-dwarfs with \kep\ magnitude ($K_p$) $> 16$} is $\sim10$. In contrast, the transiting depths of most planets discovered to date around cool dwarfs fall well below this threshold.

\begin{figure}[htb]
\center
\vspace{-10pt}
\hspace{-0.5cm}
\includegraphics[scale=0.5]{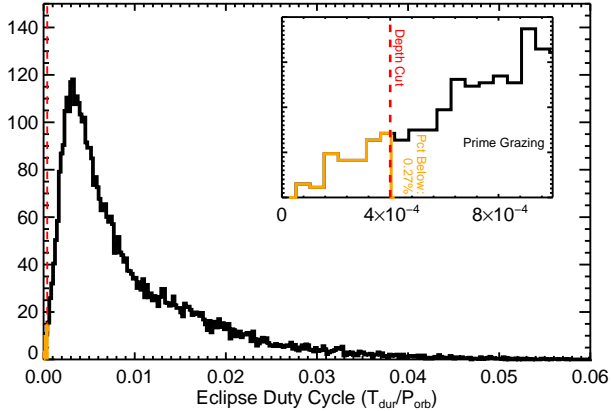}
\vspace{-0pt}
\caption{A distribution of theoretical total primary eclipse durations as fraction of orbital period for a general simulated M-dwarf EB population in the same manner as for Figure \ref{mm-ed}. Vertical axes are arbitrary units. The red dashed line marks the position of our first-round duty-cycle cut on the light curves. It lies below 99.7\% of the eclipsing cases, missing only the very grazing ones. Note that duration of ingress/egress and finite cadence would reduce the number of points expected to be captured at the theoretical maximum primary eclipse depth. }
\label{mm-dc}
\end{figure}

The eclipse duration is also an important parameter that determines the transit window and our ability to resolve the shape of the eclipse event given the cadence and time baseline of the \kep\ observations. The fraction of a light curve in eclipse, or the duty cycle, for M$+$M EBs is a function of the total radius, total mass, orbital period, and orbital eccentricity of the system. Figure \ref{mm-dc} shows the distribution of duty cycles for the same population of M$+$M EBs as in Figure \ref{mm-ed}, including grazing geometries. The orbital periods distribution is assumed to be a power law of the form $f[\log(P_{\rm orb})]\sim[\log(P_{\rm orb})]^\mu$. For the particular simulation shown in Figure \ref{mm-dc}, $\mu = 2$. The period is restricted to below 90 days, or what is feasible for an unambiguous detection and periodicity determination with at least 3 quarters of \kep\ data. All but 0.3\% of the duty cycles are greater than 0.04\%. This outcome varies little over a wide range of reasonable choices for $\mu$. 

Given the eclipse depths and durations predicted by our simulations we can devise the relevant criteria for reliably identifying EB candidates with an automated procedure. After the automated process identifies potential EBs, we then investigate the light curves by eye to identify the final sample, as we describe in further detail in the following section

\subsection{Identifying the M$+$M-EBs} \label{sec:cull}
We confine our search to all stars with at least 3 quarters of \kep\ observations (of which there are 2975 objects) such that we are uniformly sensitive to EBs with periodicities of $\lesssim$ 90 days. We use an automated algorithm to exclude the obvious non-EBs from the sample, while retaining those with suggestive features of EB candidates. We then run a semi-manual visual vetting routine on the relatively small number of targets that survive the initial vetting. Each step is described below.

\subsubsection{The Depth Cut} \label{sec:depcut}
Our automated search algorithm first checks to see if each light curve has downward flux variations that meet the depth criterion informed by our simulations. Figures \ref{mm-ed} and \ref{mm-dc} demonstrate that most M+M EBs have primary eclipse depths above 1.5\% and duty cycles above 0.04\%. For a \kep\ quarter of $\sim$90 days at $\sim$30 min cadence containing $\sim$4300 data points (neglecting intra-quarter data gaps), this amounts to requiring 2 or more of the points in the conditioned and normalized light curves to be below 0.985. Using this criterion, we end up with 150 candidates, which we subject to visual inspection.

\subsubsection{Semi-Manual Inspection in Frequency Space} \label{sec:freqcut}
For the remaining candidates, we perform a series of Fourier transforms on the un-flattened, i.e. PDCSAP, timeseries in a supervised hierarchical manner to decide whether a periodicity exists and, if so, whether it is ascribable to stars eclipsing one another. This is done through a custom user interface written in IDL. The decision to use the PDCSAP rather than the flattened light curves from the previous step is to allow the human to assess all astrophysical features holistically and to better identify the most likely cause of the feature.  

\begin{figure}[htb]
\center
\vspace{-0pt}
\hspace{-0.2cm}
\includegraphics[scale=0.5]{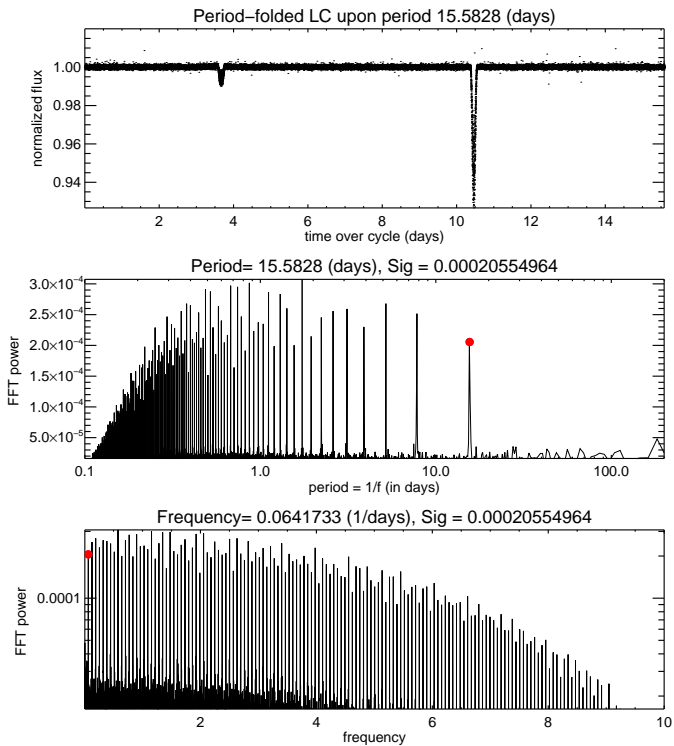}
\caption{Top panel: A sample flattened and phased light curve of the EB KID3830820. Middle panel: its Fast-Fourier Transform, plotted in terms of period = 1/frequency. The range of periods plotted is 0.1 to 200  days. The red dot, pointing to the longest-period prominent peak, is the true period of the system upon which the light curve is folded. Bottom Panel: the FFT plotted in terms of frequency. Note the slowly declining alternating peak structure, a signature of EBs. }
\label{eb-fft-phold}
\end{figure}

Interpolating the data to even spacing in time (i.e. at 29.4 min intervals), we then perform a Fast Fourier Transform (FFT), resulting in a power spectrum. When interpreted effectively, the FFT procedure can reveal much information about the causes of periodicity in a time series \citep[e.g.][]{rap14}. A light curve with no periodic variation will contain no significant peaks in the FFT. A quasi-sinusoidal signal with a wide duty cycle, as that associated with rotational modulation or BEER effects (BEaming, Ellipsoidal, Reflection -- brightness modulations associated with binaries but not their eclipses) \citep{fai-maz11}, generally shows up as a prominent and narrow fundamental peak followed by a {\emph{rapid}} decay of harmonic peaks. 
Repeated narrow features, such as the signature of an EB, results in many repeated peaks (i.e. fundamental + harmonics), perhaps alternating in height if the secondary eclipses are sufficiently prominent, with the overall height structure modulated by an envelope that is only {\emph{slowly}} declining. 
(see bottom panel of Figure \ref{eb-fft-phold}).
The central peak frequency usually does not exactly equal, but is very nearly the true periodicity or an integer fraction multiple of it. The contrast with smoothly-varying periodicities is the inherent nature of frequency transforms: narrow features in the time domain translates into broad features in the frequency domain, and {\em vice versa}. 

When significant peaks are detected in the Fourier transform, they are immediately apparent to the eye. It turns out that most of the candidates exhibit some apparent peak structure in their FFT spectrum, though most of these appear to be due to the rotational modulation of large-amplitude fast-rotators as opposed to periodic eclipses. Such objects have escaped our vetting process unscathed because their deep rotational modulation is also too rapidly varying to be fitted out without obscuring the integrity of possible eclipse signals.

As a final step, we phase-fold the light curves of potential EBs to search for the unmistakable V-shape at a very precise periodicity. In the phase-folded light curve, a non-EB might show large-scale variations akin to rotational modulation, as well as some phase-drifting, while the EB would already exhibits an obvious dip structure. Further period refinement results in a light curve akin to that displayed in the top panel of \ref{eb-fft-phold}. 

When refining the true EB periodicity for a promising target, we also perform a Box-Least Squares (BLS) search around the suspected period \citep{kov02}. This results in a more precise periodicity, though still not perfect, for the BLS was designed to find planet-transit-shaped events without obvious secondary eclipses. The final outcomes are EB orbits determined to a suitable precision for our work (i.e. typically $\sim 10^{-3}$ days), and that we have in principle found all the unambiguous EBs within some period range to which the \kep\ observation baseline ought to be sensitive. 

Running this process on our sample of M-dwarf light curves results in $17$ EB candidates with periods from 0.3 to 76 days. Since we are investigating the occurrence of unambiguously detached systems whose periods are greater than 1 day, we discard the 3 very short-period ($\lesssim 1$ day) targets, namely KID4936334, KID9077796, and KID12004834, which exhibit various degrees of contact. This is evident in the duration of their transits, which much exceed the upper limit of expected M$+$M detached EB transit durations given their orbital period. Their phase-folded light curves are shown in Figure \ref{mmeb-lcs_tight}. This criterion whittles our candidate list down to $14$.

\subsection{Further Diagnostics and the Final List} \label{sec:validation}

The EB candidates are examined further via a couple of simple diagnostics. We inspect the eclipse duration as well as centroid-flux correlations. The former test may indicate whether the eclipse could be associated with a giant. The latter validation technique is standard in identifying blends and false positives for planet detections (e.g. \citealt{bat10b, tor11}). 

The eclipse duration check involves comparing the apparent duration of the eclipse with that expected for the fiducial stellar parameters. If the actual duration much exceeds this limit, then the eclipse may be associated with a giant star or a very eccentric orbit, or is inconsistent with the detached EB interpretation. While the 3 sub-day candidates do exhibit eclipses of unusually long durations and shapes suggestive of contact (\se{freqcut}), the remaining 14 EB candidates show eclipse durations consistent with their orbital periods and estimated stellar radii.   

We also compute the positions of the brightness centroids in the pixel-level `postage stamp' data using a 2D Gaussian fit. Correlating the centroid position with flux could show photocentre shifts throughout the light curve. A significant shift during eclipse would suggest blending and light contamination, raising the possibility that the source of the light loss does not belong to the foreground target centred in the aperture. 

One candidate shows dramatic centroid shifts during eclipse. KID3830820 is blended with two other stars ($\sim 3''$ and $\sim20''$), both of which fall partially into the main aperture. The stars may be physically associated. Pixel-level light curves suggest that the eclipses are indeed associated with the main M-dwarf targeted.  

Two more candidates out of the original 17, KID5794240 and KID11548140, have survived all our search criteria but follow-up data have revealed that their eclipsing companions are non-stellar. KID 5794220, or KOI-254.01, is an M-dwarf hosting a giant planet \citep{jj12}. KID11548140, or KOI-256.01, is a post-common envelope binary whose companion is a white dwarf \citep{mui13}. 

Our final 12 M$+$M-dwarf EB candidates are listed in \Table{EBlist}. None of these stars started as Guest Observer targets. Their phase-folded light curves are displayed in \Fig{mmeb-lcs1} and \Fig{mmeb-lcs2}. Note that all the light curves show V-shaped eclipses and/or prominent secondaries, characteristic of EBs. 

\begin{table*}[htb]
\begin{center}
\caption{Final List of M-dwarf M-dwarf Detached Eclipsing Binaries in the \kep\ Field}
\label{EBlist}
\begin{tabular}{ccccccccccc}
\hline\hline
KID &  $P_{\rm orb}$  & $\delta_{prim}^*$ & $T_{dur}/P_{\rm orb}$ & $K_p$ & \# Quarters & $M_\star$ & $R_\star$ & CDPP-3 $^{**}$ & Eccentric? $^{***}$ & Notes \\
 ~  & (days) & & & (mag) & Observed & $(M_\odot)$ & $(R_\odot)$ & (ppm) & &  \\
\hline
2442084 & 49.789 & 0.251 & 0.00603 &  15.59 & 17 & 0.548 & 0.541 & 362.24 & Yes & 1 \\
3830820 & 15.583 & 0.063 & 0.0104 & 15.37 & 17 & 0.563 & 0.555 & 314.47 & Yes & 2 \\
5871918 & 12.643 & 0.256 & 0.0173 & 15.70 & 17 & 0.573 & 0.562 & 317.21 & Yes & 3 \\
6023859 & 27.010 & 0.079 & 0.00156 & 15.48 & 8 & 0.559 & 0.537 & 368.92 & Yes & 4 \\
6620003 & 3.429 & 0.033 & 0.0271 & 15.686 & 17 & 0.542 & 0.534 & 287.35 & No & 5 \\
7605600 & 3.326 & 0.193 & 0.0296 & 14.888 & 8 & 0.359 & 0.350 & 376.70 & No & 6 \\
7671594 & 1.410 & 0.106 & 0.0355 & 15.815 & 17 & 0.245 & 0.250 & 3305.8 & No & 7 \\
9772531 & 31.202 & 0.040 & 0.00319 & 15.798 & 17 & 0.462 & 0.450 & 281.53 & Yes & 8 \\
10979716 & 10.684 & 0.091 & 0.0135 & 15.774 & 17 & 0.553 & 0.524 & 268.35 & Yes & 9 \\
11546211 & 2.194 & 0.041 & 0.0301 & 15.155 & 13 (incl. Q1, Q17) & 0.339 & 0.334 & 488.50  & No & 10 \\
11853130 & 76.87 & 0.017 & 0.00135 & 15.949 & 13 (incl. Q1, Q17) & 0.407 & 0.380 & 393.58 & Yes & 11 \\
12599700 & 1.018 & 0.051 & 0.0518 & 15.780 & 17 & 0.493 & 0.480 & 1106.2 &  No & 12 \\

\hline \\
\end{tabular}
\end{center}

Notes: \\
*: $\delta_{prim}$ is the approximate depth of the primary eclipse. \\
**: CDPP-3 is the `Combined Differential Photometric Precision' over 3 hours, a metric for the noise in the light curve of the star over a typical transit timescale \citep{chr12}. Here we displayed the mean CDPP-3 over all observed quarters. \\  
***: Eccentricities are judged from the phase difference separating the primary and secondary eclipse. A phase separation of 0.5 is consistent with a circular orbit, provided the argument of periapse, $\omega$, is not exactly $\pi /2$ or $3\pi/2$. For other values of phase separation we assume the orbit to be eccentric. This includes systems for which only the primary is visible. \\ 
1: unambiguous and classic primary and secondary eclipses. Phase separation (0.181) suggests considerable eccentricity. \\
2: deeper primary accompanied by very shallow secondary with phase separation of 0.565 suggesting non-circular orbit. This star is blended with two stars (visual separation $\sim 3''$ and $\sim 20''$ whose lights contaminate that of the chief target to differing extents from quarter to quarter. As a result, this target exhibits dramatic centroid shift in eclipse. Pixel-level data suggest that the eclipses are indeed associated with the main M-dwarf target, though the depths ought to be treated with caution, since blending causes considerable variation in absolute and relative brightness from eclipse to eclipse. \\
3: clearly visible rotational modulation signature on top of the eclipse signals. The eclipses are not aligned with the rotation phase. Orbit is eccentric. \\
4: the eclipse is suspiciously sharp, i.e. short-duration, perhaps betraying the eclipse being rather marginally grazing. The true orbital period could be twice that recorded here, for visual inspection reveals the transit depths are subtlely different if folded upon $P_{\rm orb} = 2 \times 27.010=54.020$ days. In fact, RV data suggest this may be the case (Jonathan Swift, private communications). \\
5: distinct primary and secondary eclipses. Clear rotational modulation signature, appearing to be synchronized with the orbit, which has also been circularized. \\
6: notable rotational modulation in amplitude, with one periodicity being 3.9 days. Occasional but sizable flares between periods of quiescence. Orbit is consistent with circularity. \\
7: strong evidence of rotation-orbital synchronization apparent in minimal scatter of the underlying rotationally-modulated phase curve when the light curve is folded upon the orbital period. \\
8: very sharp eclipses on a very quiet star. Only primary is visible. Rounded bottom suggests the eclipse geometry is not grazing. Eclipse depth indicates the eclipsing body is $\sim 0.09 R_\sun$, which does not rule out brown dwarf or Jupiter-sized planet as the companion. If interpretation as stellar binary were robust, then the orbit must be fairly eccentric for the secondary to be out-of-view. \\
9: clearly visible rotational modulation. Orbit is not fully circularized. Effect of eclipse timing variation (ETV) is apparent, as it smears out the period-folded eclipses. A third, farther body in the system inferred from ETV has been reported \citep{bor15}. \\
10: flagged as `false positive' in MAST and catalogued as an EB in Villanova. Large-amplitude rotational modulations with periodicities 2.06 and 2.16 days, strongly suggesting rotational synchronization with the orbit.  \\
11: the longest-period EB in this sample. Only primary is visible, and it is shallow and narrow. If stellar in nature, these characteristics suggest the eclipse geometry is grazing with eccentricity. It is also not inconsistent with interpretation as a giant planet (e.g. in a most recent planet candidacy catalogue, \citet{swi15} includes this target as a cool-KOI, KOI-3263.01, with listed radius being $0.057 R_\sun$, albeit with a high false positive probability). Lucky Imaging with AstraLux at Calar Alto Observatory (PI Lillo-Box, \citep{lb12,lb14}) and recent adaptive optics with  NIRC2 at Keck Observatory (PI Ciardi) reveal a faint visual companion at $\sim$0.8". Followup observations are continually updated through the \kep\ Community Follow-up Observation Program, CFOP (https://cfop.ipac.caltech.edu). \\
12: a restless and regularly flaring system. Notable rotational modulation synchronized with eclipses.  \\

\vspace{20pt}
\end{table*}

The longest orbital period is 76 days, or 0.21 years (log $P_{\rm orb}$ = -0.68), and many are below 3 days, or 0.008 years (log $P_{\rm orb}$ = -2.1). The fact that we have few detections at long periods and many in close orbits is consistent with expectation despite ignorance of the actual binarity distribution, for the average geometric transit probability declines rapidly with orbital period, and the growing likelihood of non-zero eccentricities at larger orbital separations make it less likely to observe both the primary and secondary. Hence we would expect fewer wide-orbit EBs to be detected. This expectation is an observational bias that relates to the fact that longer period EB detections have the greatest influence in our conclusions about overall binarity rates, for they carry the greatest statistcal weight, being the least likely to be found. We discuss the statistical framework from which we draw inferences in \se{EBstats}.

\subsection{Comparison to existing \kep\ EB catalogues} \label{sec:comparelist}

EBs in \kep\ have been tracked and studied by previous groups. \citet{cou11} catalogued and modelled the light curves of all detached EBs with primary T $< 5500$K and orbital periods under 32 days in campaign 1 of \kep. The \kep\ Eclipsing Binary Working Group have produced the Villanova catalogue, a running compilation of all detected EBs in the \kep\ dataset (for methods and a description of the catalogue, see \citealt{prs11, sla11, mat12}). The catalogue contains classification of each candidate,
providing a benchmark against which we can compare the outcome of our search procedure, and to cross-check their catalogue completeness for M$+$M EBs. Within the DC13 sample, there are 19 overlaps with the Villanova Catalogue, which includes all the \citet{cou11} EBs in the same mass range. All seventeen of our original EB candidates are among this overlap. Of which three (KID4936334, KID9077796, and KID12004834) satisfy our two initial cuts but possess orbital periods within 0.4 days and unusually large duty cycles, making their interpretation as normal detached binaries unreliable. They have therefore been excluded and we have limited the scope of our study to periods above 1 day. These very short-period binaries are shown in figure \ref{mmeb-lcs_tight}. Yet another two `EBs', KID5820218  and KID9892651, are unconvincing upon visual inspection. Figure \ref{mmeb-lcs_missed} shows their respective light curves folded upon their documented periods. The eclipses, if actually present and due to stellar sources, are very low signal-to-noise and would represent either a marginally grazing geometry or a nearly substellar eclipsing companion, or both. 
 
We have encountered a few other unambiguously detached M-dwarf EBs not documented in the Villanova Catalogue, namely KID5769943 ($P_{\rm orb}$ = 1.01~d), KID7938883 ($P_{\rm orb}$ = 0.490~d), KID8949316 ($P_{\rm orb}$ = 0.604~d), and KID12009213 ($P_{\rm orb}$ = 19.4~d). However, they are all part of the Guest Observer (GO) program GO20001\footnote{Interestingly, GO20001 contains 1196 targets and was intended to identify new short-period M-dwarf eclipsing binaries for stellar characterization.}, whose targets were observed for one quarter only (between Q6 and Q9), hence do not fulfill the criteria of having an observing baseline of at least three quarters.    

\subsection{Statistical Analysis and the NSPS} \label{sec:EBstats}

Since \kep's launch, a number of statistical studies have been published to infer the occurrence and distribution of planets around their host stars, studying dependencies on planet radius, mass, orbital period, eccentricity, planetary multiplicity, and orbital alignment \citep[][hereafter MS14]{how12, dc13, pet13, fre13, fm14, bj15, ms14}. Most of these studies provide a measure the number of planets per star, also known as NPPS (see, e.g., MS14).  \citet{you11} provides a generalized statistical framework for measuring NPPS using a maximum likelihood analysis assuming Poisson statistics. The non-parametric version proceeds as follows to give a point-estimate of companion occurrence integrated over the entire phase space of system properties searched. Referring to the definition of NPPS in their Equation~9: 
\be
f_l = \frac{N_l}{\eta_l N_\star},
\label{you11-9}
\ee
where $N_\star$ is the total number of stars in the survey, $l$ labels one of the multiple bins into which detections are divided (e.g. a range of orbital periods), $N_l$ is the number of planet detections in bin $l$, and $\eta_l$ is the detection efficiency in that bin. In effect, $1/\eta_l$ serves as a `statistical weight' on the fraction of planets detected within the survey star sample, which corrects the actual detections to the true number that exists inside the bin.  Summing $f_l$ over all bins give the overall planet occurrence: 
\be
{\rm NPPS} = \sum_l f_l.
\label{you11-nppstot}
\ee


The problem of stellar multiplicity far predates that of planet multiplicity (see \se{intro}), but the techniques used to measure planet occurrence from a transit survey are directly analogous to our goal of measuring binarity from stellar eclipses. The above methods constitute the starting point of our analysis which, in many respects, is a simpler problem than exoplanet statistics; detection of an object as large as a star `transiting' a star is not hampered by low signal-to-noise, unlike detecting planet transits. Also, since we are concerned only with M$+$M dwarf binaries, several simplifications can be made. As has been demonstrated in \se{depcut}, all but the most grazing of eclipsing geometries ought to produce signals that far exceed the amplitude of, and have very different shapes than  typical stellar photometric variability. Thus, false positive rates should be low for our study and the discovery efficiency is chiefly limited by the geometric eclipse probability.

In general, the eclipse probability for binary stars is:  
\be
P_{\rm tr}(R_{\rm tot}, a, e, w) =   {\rm Max}\left[\left(\frac{ R_{\rm tot}}{a}\right) \left(\frac{1\pm e \sin \omega}{1-
e^2}\right)\right], 
\label{eb-trans-prob}
\ee

\noindent where $R_{\rm tot} \equiv R_{\star1}+R_{\star2}$, $e$ is the orbital eccentricity, $\omega$ is the argument of periastron, and $a$ is the semi-major axis, related to the orbital period by Kepler's 3rd Law. The `$+$' in the $\pm$ refers to the primary, whereas the `$-$' applies to the secondary eclipse. Marginalizing over all $\omega \in [0, 2\pi]$ results in
%

%
\be
P_{\rm tr}(R_\star,M_\star,P_{\rm orb}) = R_{\rm tot}\left[\frac{4\pi^2}{P_{\rm orb}^2 G M_{\rm tot}}\right]^{1/3} \left(\frac{1}{1-\langle e^2 \rangle}\right),
\label{eb-trans-prob-porb}
\ee

\noindent where  $M_{\rm tot} \equiv M_{\star1}+M_{\star2}$  and $\langle e^2 \rangle$ is the average squared eccentricity of the population of binaries. For main-sequence stars, $R_\star \approx M_\star$, in Solar units, and as a result $R_{\rm tot} \approx M_{\rm tot}$. Eliminating constants, we may write the following approximate dependency: 
\be 
P_{\rm tr}(M_{\rm tot},P_{\rm orb}) \sim \left[\frac{M_{\rm tot}}{P_{\rm orb}}\right]^{2/3}\left(\frac{1}{1-\langle e^2 \rangle}\right). 
\label{eb-trans-prob-prob-2}
\ee
More massive systems with shorter orbital periods are more likely to transit. A population with a higher average eccentricity is more likely to eclipse. The distinction with exoplanet transits is that both components of the binary system contribute. $P_{tr}$ is taken as the detection efficiency $\eta_l$, where bin $l$ is identified by the system orbital period. 

When dealing with a large sample of stars, the probability of transit around every star must be accounted for. In practice, this condition modifies $\eta$ for each detection. As MS14 puts it, $\eta$ should be a factor that obeys the following statement: `if a very large number of {\emph{stars}} identical to {\emph{star} i} were distributed randomly around all the stars in the survey, only a fraction $\eta_i$ could have been detected.' We treat this issue by essentially averaging over the theoretical $\eta$ around each primary target star in the survey sample. 

In the small-bin-limit, we can compute the {\emph{Number of Stars Per Star}}, NSPS$(M_\star, P_{\rm orb})$ using the corresponding $\eta(M_\star, P_{\rm orb})$ to estimate the number of undetected systems based on each detection. Binning the results into ranges in parameter space of interest, be it in $M_\star$ or $P_{\rm orb}$, we may obtain the distribution function in these chosen parameters. This would be the case if the system configurations are perfectly known.  

However, an eclipse light curve alone, unsupplemented by spectroscopic information, encodes a limited amount of information. While orbital periods can be measured very accurately, it is difficult to estimate the primary and secondary masses of each given system. So while vast simplifications occur for EB detections, the non-negligible size and self-luminous nature of both components involved introduces some additional complications. To compute $\eta(M_\star, R_\star, P_{\rm orb})$ in a meaningful way for each detection, we use a Monte Carlo approach. Each target in our sample, be it single or multiple, is attached to a mass and a radius value derived from DC13's isochrone fits. The primary mass $M_{\star1}$ and radius $R_{\star1}$ are equated with these values. This is a reasonable assumption because, for equal-mass systems, the photometry of the system is not too dissimlar to each individual star and, for low mass-ratios, flux from the primary will dominate the system flux.
  
We draw the the secondary star's mass, $M_{\star2}$, from a mass ratio distribution of form $f(q) \sim q$, where the radius is calculated using the empirical mass--radius relationship from \citealt{boy12}. 
Next, we draw the orbital eccentricity ($e$) from an eccentricity probability distribution and the argument of periastron ($\omega$) from a uniform distribution, and the inclination ($i$) from a distribution uniform in cos($i$). 

Unfortunately, the true eccentricity distribution for close M$+$M binaries is not well measured. An exploration of this issue by an extensive RV survey for low-mass spectroscopic binaries \citep{may00, udr00} show that binaries with components on or beyond the main-sequence and periods $< 10$~days have been circularized, while for periods longer than 10 days the eccentricity distribution is a steeply rising function of orbital period. \citet{rag10}'s study of solar-type stars report similar findings, namely orbital circularization for binaries under 12-day periods. Beyond this threshold, any eccentricity up to 0.6 is equally likely. \citet{hal03} observe that, for FGK binaries, those with higher mass ratios preferentially possess dampened eccentricities relative to their more unequal mass ratio counterparts. These findings are consistent with tidal evolution theory. The tidal circularization timescale, which governs how likely a binary system is found circularized, is a rising function of orbital period and mass ratio \citep{hut81}.  

Despite these clues, no functional forms exist for a joint probability distribution of eccentricity in terms of orbital period and mass ratio. The marginal cases in our own EB sample, namely two systems with $P_{\rm orb} \sim 10$ days and $P_{\rm orb} \sim 12$ days (refer to \Table{EBlist}), are both eccentric, as evidenced by the timing of secondary eclipse relative to the primary eclipse. Therefore, we adopt a piecewise eccentricity distribution, where $e\,(P_{\rm orb} < 10 ~{\rm days}) = 0$ and $e\,(P_{\rm orb} > 10~{\rm days}) \in [0.0, 0.6]$ uniformly. We also perturb the upper bound between 0.4 and 0.8 to test the robustness of the final statistic to this assumption.    

The process for simulating orbits for each EB detection is repeated 1000 times 
around every star surveyed, yielding a distribution in detection efficiency $\eta$ at every detected EB orbital period, $\bar{\eta}(P)$. The detection is then weighted by $\eta^{-1}$ and normalized by the total number of survey stars to reach a distribution of $f_l$ as defined in Equation (\ref{you11-9}). From here, we can give our best point estimate of NSPS via eqn (\ref{you11-nppstot}). In the small-bin limit where each detection $i$ constitutes its own bin, NSPS = $\sum_i f_i$.

Since the actual number of detections is quite small, to reach a period {\emph{distribution}} of the binaries via the typical route of constructing a weighted histogram is problematic. In particular, binning in histograms is arbitrary and discontinuous, especially if each bin contains only a handful of detections. A subtle change in the binning scheme could drastically change the shape of the distribution. 

To avoid this issue, we adopt the maximum likelihood analysis for estimating a parametric distribution of companion occurrence following the procedure detailed in \citet{you11}. So long as the target distribution can be reasonably parametrized, this approach does not involve binning, and also yields confidence intervals. We begin by writing a general differential distribution: 
\be 
\frac{\partial f(x)}{\partial x} \equiv Cg_\alpha(x)
\label{you11-eqn14}
\ee 
so that $df = (\partial{f}/\partial{x})dx$ represents the probability of a primary M-dwarf star having a binary companion in the interval $dx$. $x$ is some general property of the system and $\alpha$ is a shape parameter that defines the functional form of the shape function, $g$. $C$ is an amplitude or normalization factor.

We set $x = \log_{10} P$ and $\alpha$ to be a power law exponent that parametrizes the underlying period distribution, such that: 
\be 
g_\alpha(x) = e^{\alpha x} = \left(\frac{P}{P_0} \right)^\alpha
\label{you11-powerlaw}
\ee 
here $P_0$ is a reference period value, which is set to 1 day.


The total number of binary companions around $N_\star$ surveyed stars is then: 
\be 
N_{tot}=C~\sum^{N_\star}_{j} \int{g_\alpha(x) dx} 
\label{you11-ntot}
\ee 
as in eqn (15a) of \citet{you11}. 

We can define a `shape integral', $F$, analogous to eqn (17) of \citet{you11}, to weigh the shape function over all the stars surveyed and the detection efficiency:
\be
F_\alpha \equiv \frac{1}{N_\star} \sum^{N_\star}_j\int{\eta_j(x)g_\alpha(x)dx}
= \int{\bar{\eta}(x)g_\alpha(x)}dx
\label{you11-shapeintegral}
\ee 
where  
\be 
\bar{\eta}(x) = \frac{1}{N_\star} \sum^{N_\star}_j \eta_j(x) 
\label{etabar}
\ee 
this is essentially what we mean by computing $\bar{\eta}(P)$ earlier via Monte Carlo simulation. The expected total number of binary detections, given $g_\alpha(x)$, is $N_{\rm ex} = N_\star C F_\alpha$. $\eta_j$ is the net detection efficiency of a companion around star $j$ with system property $x$.  

The likelihood of the observed data, i.e. number of detected EBs, $N_{\rm eb}$, is a product of their Poisson likelihoods: 
\be 
\tilde{L}= \left[\prod^{N_{\rm eb}}_{i=1} df_i \right] \exp(-N_{\rm ex}) 
\ee
which, given the formalism introduced thus far, can be rewritten as:
\be 
L = \left[C^{N_{\rm eb}} \prod^{N_{\rm eb}}_{i=1} g_\alpha(x_i)\right] \exp(-N_{\rm ex}) 
\label{you11-eqn19}
\ee
with the planet parameter differential, $dx$, eliminated since it represents a constant interval. It turns out that the logarithm of the likelihood in eqn \ref{you11-eqn19} can be put in the following form (see \citealt{you11} and references therein), the sum being performed over all EB detections: 

\be
\begin{split}
\ln{L_\alpha} &= -N_{\rm eb}\ln F_\alpha \\
&+ \sum^{N_{\rm eb}}_i \ln(g_\alpha(x_i)) + N_{\rm eb} \left[\ln \left(\frac{N_{\rm eb}}{N_\star}\right) - 1 \right]  
\end{split}
\label{you11-logL}
\ee 

\noindent the best-fit normalization, C, is the one that maximizes likelihood, i.e. $\partial \ln L / \partial C = 0$ and is precisely: 
\be 
C_\alpha = \frac{N_{\rm eb}}{N_\star F_\alpha}
\label{you11-C}
\ee 
i.e. the most likely C is that which matches the expected number of detection, $N_{\rm ex}$ and that actually detected, $N_{\rm eb}$.    

Our goal is to find the best-fit $\alpha$ and its confidence intervals. We do so by searching over a grid of $\alpha$'s and computing the log-likelihoods of the data as given in eqn \ref{you11-logL}, calculating $g_\alpha$ and $F_\alpha$ for each. Contours in log-likelihood can be compared to a Gaussian distribution. Around the maximum likelihood, $\ln(L_{\rm max})$, the $n\sigma$ confidence interval involves $\alpha$ values that results in $\ln(L_\alpha) \geqslant \ln(L_{\rm max}) - n/2$. The NSPS is given by ${\rm NSPS} = N_{\rm tot} / N_\star$, $N_{\rm tot}$ from eqn (\ref{you11-ntot}). The best-fit value is always consistent with that derived from eqn (\ref{you11-nppstot}).




~


\section{Results and Discussion}\label{sec:discuss}

\begin{figure}[htb]
\center
\vspace{-10pt}
\hspace{-0.5cm}
\includegraphics[scale=0.5]{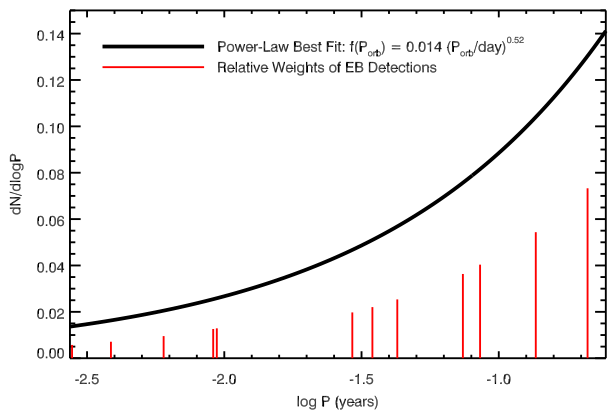}
\vspace{-10pt}
\caption{The inferred power-law orbital period distribution and overall occurrence of M$+$M binaries in the \kep\ field from EB detections. The best-fit power-law exponent, $\alpha$, and its $1\sigma$ interval, is $0.52\pm0.22$. This corresponds to NSPS $= 0.106 \pm0.02$. The red vertical lines delineate the actual EB detections and their relative weights due to detection efficiencies.}  
\label{ebstats-q=1}
\end{figure}

\begin{figure*}[htb]
\center
\vspace{-10pt}
\hspace{-0.5cm}
\includegraphics[scale=1.0]{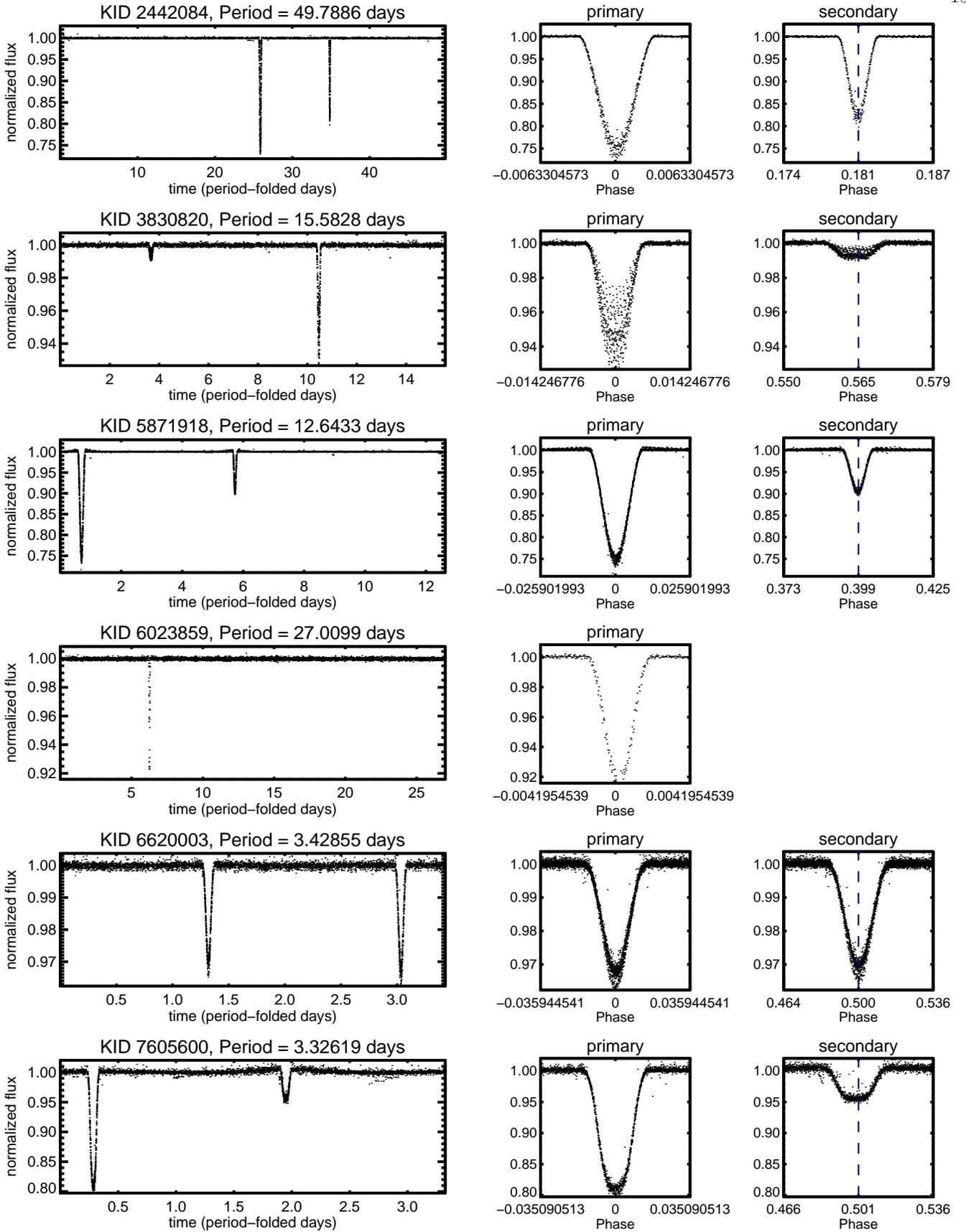}
\vspace{-0pt}
\caption{The period-folded light curves of M$+$M eclipsing binary targets that made our final list. Note the dramatic in-eclipse scatter in the light curve of KID3830820, which is chiefly due to varying degrees of blended light with neighbours from quarter to quarter. See Table \ref{EBlist} for descriptions of each pre-flattened light curve.}
\label{mmeb-lcs1}
\end{figure*}

\begin{figure*}[htb]
\center
\vspace{-10pt}
\hspace{-0.5cm}
\includegraphics[scale=1.0]{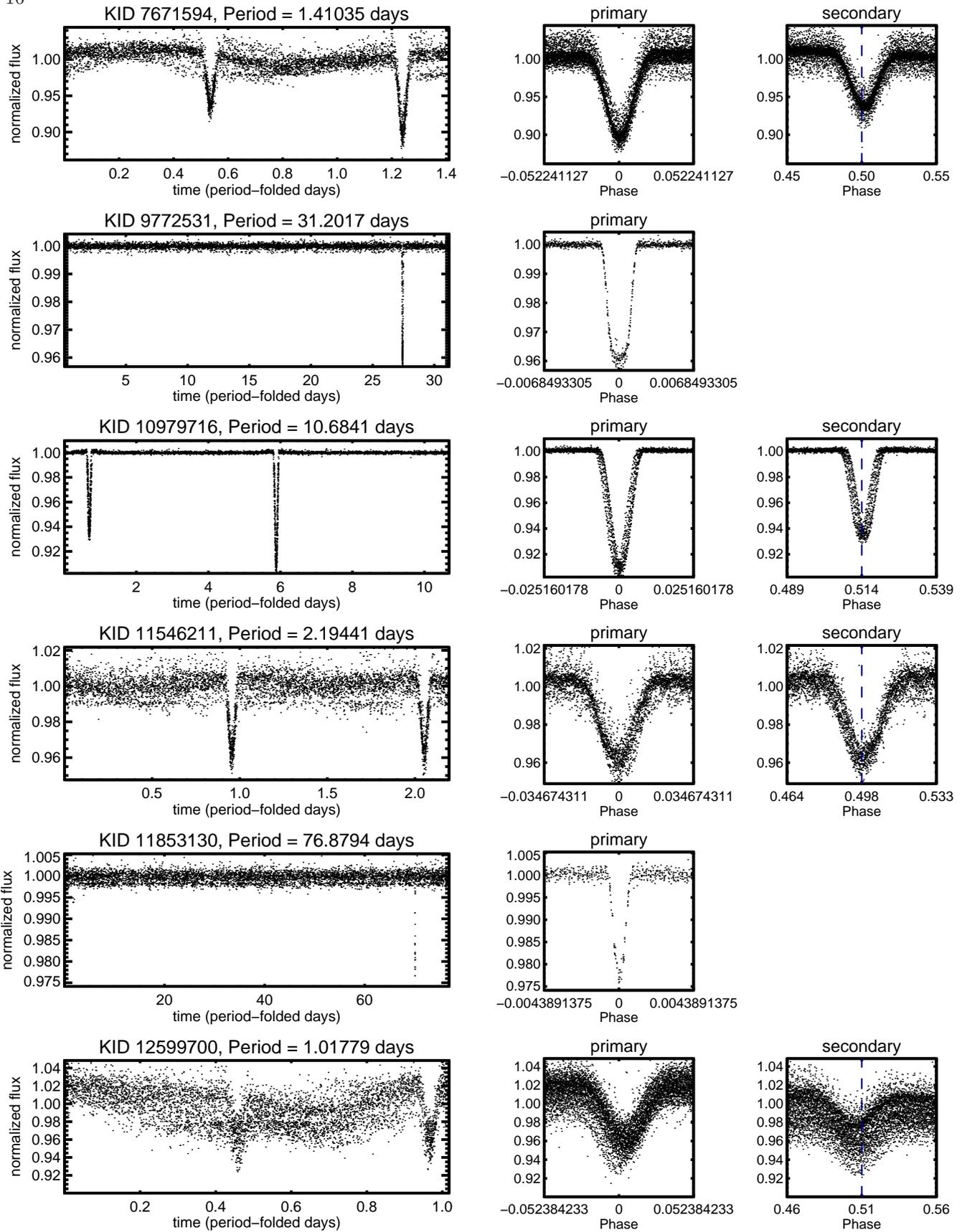}
\vspace{-0pt}
\caption{Period-folded M$+$M EB light curves continued. See Table \ref{EBlist} for descriptions of each pre-flattened light curve.}
\label{mmeb-lcs2}
\end{figure*}

\begin{figure*}[htb]
\center
\vspace{-10pt}
\hspace{-0.5cm}
\includegraphics[scale=1.0]{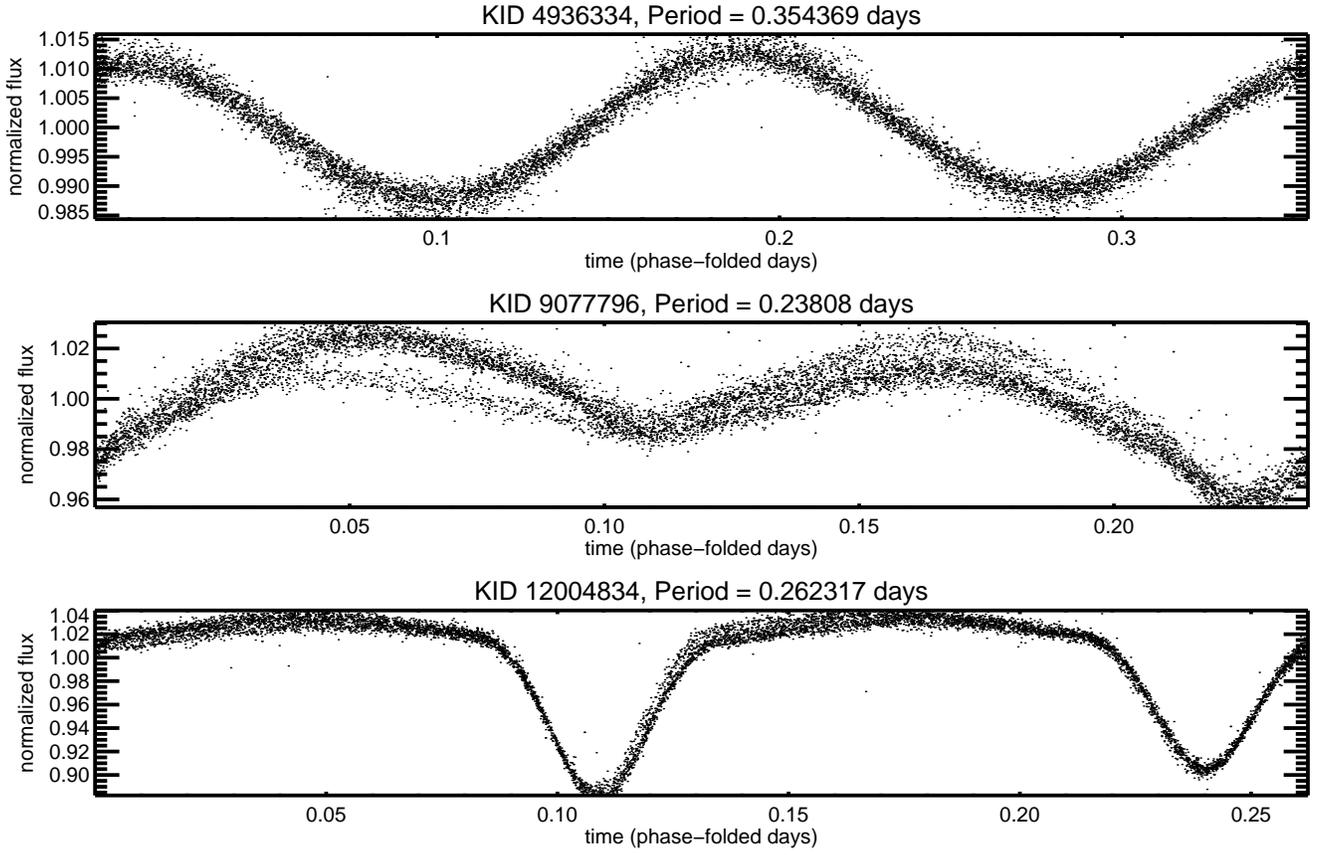}
\vspace{-11.5cm}
\caption{Three very tight EB candidates 
which have duty cycles inconsistent with their theoretical maximum. They exhibit some level of contact. KID4936334 has anomalous photometry -- its DC13 catalogue entry registers $M_\star = 0.09 M_\sun$, $T_{\rm eff} = 3433$, and $Z = -2.0$. These candidates have been discarded from the final list. All three are Villanova objects.}
\label{mmeb-lcs_tight}
\end{figure*}

\begin{figure*}[htb]
\center
\vspace{+10pt}
\hspace{-0.5cm}
\includegraphics[scale=1.0]{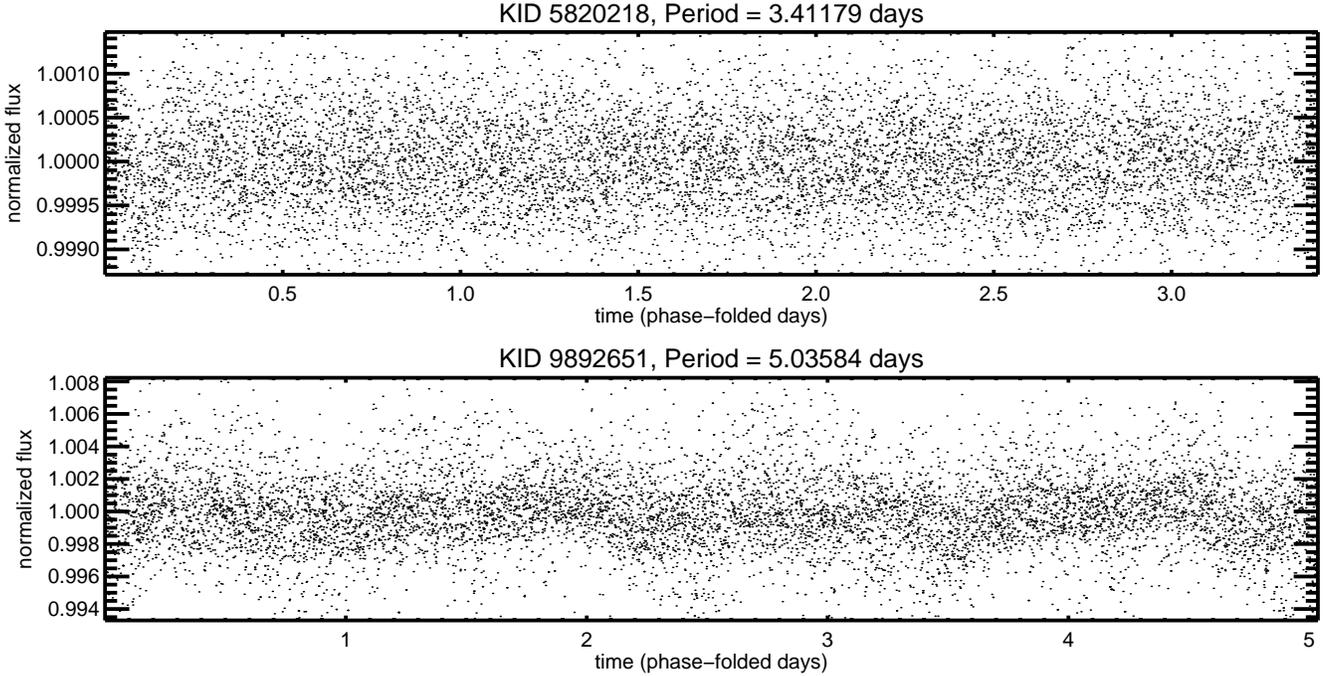}
\vspace{-13.5cm}
\caption{Two unconvincing Villanova EBs. They do not meet the criteria for detection by our algorithm, hence are not included in our analysis. }
\label{mmeb-lcs_missed}
\end{figure*}

We have selected M-dwarfs from the \kep\ Mission with more than 3 quarters of photometry and identified the eclipsing binary systems among them. These constitute all the M$+$M close binaries with favourable orbital inclination geometries to result in detection as EBs. A statistical framework akin to that used for exoplanetary occurrence rates is employed to analyze the data and infer the orbital period distribution and overall close binarity rate of the detached M-dwarfs in the \kep\ field. 
We summarize the result of the exercise described thus far with Figure \ref{ebstats-q=1}. Integrating our derived power-law distribution over the period range spanning 1--90~days results in NSPS$= 0.11\pm0.02$. Perturbing the eccentricity distribution by setting the maximum eccentricity to 0.4 and 0.8 result in a systematic 0.01 upward and downward adjustment to the median, respectively. 


\subsection{Previous Studies}\label{sec:prevstudies}

The most directly comparable published studies of M-dwarf multiplicity sensitive to this tight period range are FM92 and CBK12, both of which use RV variation overtime as an indicator of stellar companionship within a few AU. In the case of FM92, the typical dataset for a primary target contains 15 spectra over 4 years. In principle, the companion sensitivity is well below the substellar threshold. They conclude that the binarity fraction within 0.4 AU is $1.8\pm1.8\%$. CBK12's study is based on RV variability measured from stellar spectra in the Sloan Digital Sky Survey (SDSS) over 2-30 days. They infer a binarity fraction in a comparable orbital distance to be $3-4\%$. The present study uses photometric time series data and the detection of eclipsing binaries to argue a binarity fraction of $11\%\pm2\%$ within a 90-day period. 
Both previous results are in tension with the present study.

In what follows we examine the particulars in the previous binarity studies, from sample selection to analysis technique, to highlight the merits and limitations of each, and to assess the appropriateness of direct comparison.

\subsubsection{FM92}\label{sec:FM92}
The FM92 sample consists of stars targeted as part of an RV program to search for substellar companions around mid- and late-M-dwarfs \citep{mb89}. As such, the affiliated spectral identifications of their objects are all M2 or later with V$<$11.5 mag. To accentuate the dynamical signature of brown dwarfs, which is the original objective of this observing program, the FM92 RV sample excluded targets that possess a known companion within $10''$ to avoid two stars falling on the spectrometer entrance slit. Since targets amenable to RV studies must be very bright, the final sample in FM92 consists of 62 M-dwarfs in a magnitude-limited sample all residing within $\sim15$pc of the Sun, at declinations of $-10^\circ$ to $+50^\circ$. The authors assess completeness factors in terms of orbital elements and companion mass, and adopt a conservative estimate of this factor to derive the total number of companions. Among these M-dwarf primaries, only a single companion was detected within 0.4~AU. Thus, the measured binarity in the semi-major axis bin 0.04 - 0.4~AU is drawn from this sole example. 

FM92 presumably use the central limit of the Poisson distribution to arrive at their binary rate of $1.8 \pm 1.8$\%. Note that, with only a single detection, it would be more appropriate to use  binomial statistics, which gives rise to an asymmetric probability distribution of $2.8_{-1.6}^{+2.2}$\%, i.e. with a larger tail towards higher values.

\subsubsection{CBK12}\label{sec:CBK12}

CBK12 use M-dwarf spectra collected from SDSS. The sample is nominally magnitude-limited (the initial selection criterion enforces $16<i<20.5$), and in total contains 39,543 M-dwarfs (though in practice, the sample used for companionship search is much smaller, see below). 

To contrast with the FM92 targets, whose spectra collection is more systematic, the SDSS spectra are recorded somewhat haphazardly over very short time baselines and the sampling is relatively sparse. The intent of SDSS spectra is for classification and characterization as opposed to monitoring for radial velocity variations. In other words, it is a serendipitous RV survey.

In their analysis, CBK12 used a combined spectrum formed by co-adding all exposures for each star. Typically they are three 15-minute exposures taken sequentially, such that the time baseline spanned lies within an hour. In rare circumstances, though, the set of spectra used to construct the combined spectra would be separated by baselines as long as days, lending themselves to the possibility of observing radial velocity variations. Within the M-dwarf sample, 1452 stars have observations spanning over 2-30 days. It is within this M-dwarf subsample that the authors searched for RV variations indicative of companions. 

The SDSS M-dwarfs are large in number and span a relatively large volume in space. The stars are sampled in small patches that are uniformly distributed in the northern sky ($\delta > -20^\circ$) \citep{sto02,aba09,yan09}. 
However, CBK12 is not volume-limited and not really magnitude-limited, either. Referring to Figure 1 therein, the actual magnitude distribution of the sample is unimodally peaked at $i\approx16.8$ and tapers to nearly none at $i\sim18$. 


Many SDSS stellar spectra have S/N $>$ 30, sensitive to radial velocities down to $\sim 4$km/s, making them amenable to identifying bulk kinematic properties of galactic structure, such as stellar streams \citep{yan09}. However, they are not ideally set up for RV work. The binary detections are identified by assuming all detected RV variability above a threshold can be attributed to the influence of a binary companion. The details of the criterion itself assumes an underlying and stationary error distribution of RV variability, which is inferred from a control sample of targets whose RV time baselines are $\lesssim 2$ hours. Applying this criterion, there are 22 detections throughout the CBK12 sample. After completeness corrections, they conclude a close binary fraction of either 3\% or 4\% within 0.4 AU, depending on their assumption of the frequency distribution in semi-major axis. The Bayesian posterior $1\sigma$ confidence interval is $\sim1\%$ for each scenario.

\subsection{Particular Points of Comparison}\label{sec:verdict}

\subsubsection{Sample Comparability}\label{sec:sample comparability}

Are we really all probing the same population of objects? 
As mentioned under \se{FM92}, the primary stars in the FM92 RV sample are predominantly of spectral type M2 and later. In contrast to the typical \kep\ M-dwarf target, which lies $\sim200$pc away, the FM92 stars reside in the very local solar neighbourhood ($<$15pc). In addition, all potential targets harbouring a known companion within $10''$ have been excluded. It is now well-known that spectroscopic and contact binaries are likely to have distant triple companions \citep{tok06, pr06}, whereas FM92 assumes this correlation to be negligible. Hence the cut made by FM92 could bias their sample in the search for close-in companions. The FM92 sample is magnitude-limited ($V<11.5$), though the targets are ultimately chosen for their amenability to substellar companion RV searches. There are 62 primaries in total. 

For CBK12, the parent spectroscopic M-dwarf sample is extracted from a deep magnitude-limited all-sky survey, resulting in a large number of M-dwarf targets with medium-resolution optical spectra ($R \sim 1800, \lambda < 920nm$). The final subsample of 1452 candidates, deemed appropriate for a companion search, is a somewhat serendipitous selection from the overall survey (see \se{CBK12}), whose brightness distribution is unimodal and peaks at $i\sim16.8$. Though the DC13 sample satisfies the colour cuts explicitly imposed by CBK12 (i.e. $i-z > 0.2$, $r-i > 0.5$), as a whole, the CBK12 sample is redder than DC13. CBK12 has median $i-z \gtrsim 0.5$, mode $\sim 0.55$ (see Fig 1d therein), compared to median $i-z \sim 0.4$ for DC13, the mode being $\sim 0.35$. This hints that CBK12 samples a somewhat later M-dwarf population than our study. 

A known limitation associated with SDSS spectroscopy is its fiber collision limit. Two objects within $40''$ could fall within the same fiber, and only one of them would have its spectrum taken. If resolved tertiary companions are commonplace around tight binaries, then only a fraction of every such system surveyed would attain a spectroscopic measurement of the inner binary and detect an RV variation (Cullen Blake, private communications). The consequence of missing such systems would be an underestimated binarity rate, an effect to which CBK12 may be subjected.          

Relative to these former works, the present study claims no superiority in the definition of its original sample selection. As did its predecessors, it makes use of an existing dataset whose initial purpose was not necessarily to be conducive to rigorous statistical analysis of stellar multiplicity. It is neither volume- nor magnitude-limited (see Figure 
\ref{dc-sloanr} for magnitude distribution of sample stars). 
The number of stars monitored by \kep\ -- $>3000$ within $\sim200$pc-- is an improvement over both, and samples from both nearby and somewhat more distant stellar environments., though the corrrection factors are large for longer periods. Note that the \kep\ M-dwarf targets are dominated by early-types.               

The differing populations studied by each study (early- vs mid- vs late-M-dwarfs) may contribute some intrinsic, physical differences between each measure. After all, multiplicity is a heavily mass-dependent phenomenon. However, here one expects the difference to be small. Even down to the very-low-mass (VLM) and substellar regime, the overall multiplicity statistics are within a factor of 2 of their low-mass star counterparts \citep{dk13}. In addition, though multiplicity is generally a decreasing function of mass, one would expect from scaling and observational evidence that the orbital period distribution of companions also moves inwards, i.e. the close companion fraction would inflate for the VLM stars relative to more massive objects. The two effects work in opposing directions.

\subsubsection{Potential Sample Biases}

One concern in surveys with magnitude limits, however implicit, is that they tend to favour the inclusion of unresolved binaries due to their greater combined brightness. In other words, the close binaries are surveyed out to a larger volume than the single stars, resulting in an overestimate of their numbers relative to a volume-limited survey. This pitfall is known as the `Branch Bias' \citep{bra76}. To counteract this bias, a correction factor that is dependent on the flux ratio distribution of the binary components may be applied to convert into a volume-limited measurement, whose analytic form is given in \citet{bur03} (eqn (4) and (5) therein). In the case of a strictly magnitude-limited imaging survey and assuming all binaries consist of equal-mass components, this issue may produce up to $2.5\times$ upward bias in the binarity fraction measured (see, e.g., \citet{bur03, jod13}). The reality is usually less extreme, though still appreciable. For a realistic binary population with {\emph{overall}} unresolved binary fraction of 20\% - 40\%, the overestimate inferred from the EB sub-population in a magnitude-limited survey can be $\sim$20 - 60\%. Scaling the observed NSPS for our period range of interest by this factor reduces the best estimate to $\sim$0.07 - 0.09, which is less discrepant from the previous measurements. Adding this possible systematic bias in quadrature with the formal error on the fit, we adopt -0.04 as the lower uncertainty on our NSPS. Note that FM92 make an effort to account for the Branch bias by discarding 3 SB systems whose individual components do not meet the magnitude selection criterion. CBK12 have not attempted to correct for this bias. GAIA parallaxes will be authoritative in the resolution of this bias.  

The unresolved nature of these binaries might pose another selection bias for the sample -- they may distort the colour, hence affect their initial selection and identification in DC13, which uses broadband photometry to constrain the stellar properties. DC13 in turn makes a temperature cut at $T_{\rm eff} < 4000K$ and a surface gravity cut at $\log(g) > 3.6$ to restrict their target range to low mass dwarf primaries. Bias in the sample can occur when M-dwarf binaries tend to be systematically misclassified due to its colour distortion and, as a result, are ejected from the sample to begin with. Of course, it is also possible that binarity in non-M-dwarf primaries may effect a colour change that causes them to be preferentially mistakened as M-dwarfs. We deem both effects to be small. On the lower main sequence, luminosity scales steeply with mass -- as approximately $M_\star^5$. Equal-mass binaries should present minimal colour difference from the primary itself since the primary and secondary would present nearly identical spectral energy distributions (SED). Unequal-mass binaries should expect the primary SED to overwhelm the contribution to the light detected overall. 

Though the DC13 stellar parameters are considered to be improved upon the original KIC, recent studies suggest a possible systematic underestimation in their radii \citep[e.g.][]{eve13,pla14,new15}, by $\sim$15\%. If the true radii of the M-dwarf primaries are larger than assumed for this analysis, greater detection probabilities are implied, and the weight for each detection decreases by the same factor. Such an error could contribute a $\sim$15\% overestimate in the NSPS.   

~
\subsubsection{Completeness Estimates}\label{sec:completeness estimates}

Both former studies have strived to make conservative completeness estimates. This means their surveys are probably more sensitive than assumed for their occurrence rate calculations, rendering the results upper limits.  

In this work, we have assumed that all eclipsing systems, even grazing ones, would be easily detected. This is a good assumption for stellar eclipses until the extremal grazing regime (see \Fig{mm-ed}), hence we may be overoptimistic in our detection sensitivity in our formal calculations. An overestimate in completeness translates into an underestimated multiplicity fraction. If this is the only factor biasing the completeness, our result would actually be a lower bound. 

In the opposite direction, we could be `over-complete'. That is, perhaps we have mistook a few systems for genuine EBs. It is not far-fetched to expound that our `EB detections', especially the relatively shallow ones and those for which the secondary eclipse are invisible, may in reality be due to the transits of low luminosity companions (e.g. brown dwarfs and giant planets). In effect, we may worry about the `false-positive rate'.

Existing knowledge of brown dwarf and giant planet occurrences as companions to M-dwarfs is that they are very rare \citep[e.g.][]{mb89,end06,kra08,det12, dc13, ms14, bow15}. According to equation (8) of \citet{jj10} which describes the fraction of stars hosting giant planets as a function of stellar properties, the M-dwarf targets deemed to be EBs in our sample have less than 4\% probability of hosting a giant planet within 2.5 AU. Note that one such false positive in the original EB candidate list (KID5794240, or KOI-254.01) has already been identified, confirmed, and excluded. The chance that another giant planet might mimic an M-dwarf EB would push the giant planet host fraction among these M-dwarfs to $2/14 = 14\%$ within $<0.4$ AU, which is highly unlikely. RV and imaging campaigns have generally found the occurrence of brown dwarf companions to stars regardless of spectral type and orbital separation to be $\sim0-5\%$ \citep{mb89, kra08, det12, bow15}, comprising the so-called `brown-dwarf desert'. Among the \kep\ low-mass stellar targets, only the brown dwarf LHS6343 \citep{jj11, mon15} has been found to eclipse an M-dwarf (not included in the DC13 catalogue). Still, the likelihood of finding transiting BDs and Jupiters in the remaining sample should be very small and thus unlikely to affect our results, and future RV follow-up of these binary systems will best provide confirmation of this assumption.

A couple of systems in this sample do present a shallow primary and lack a secondary eclipse (KID9772531, KID11853130). Both have relatively lengthy periods. Throughout this analysis, we have been interpreting their nature as systems with eccentricity. However, it is also possible that one or both of these are false-positives, for their eclipse depths suggest companion radii consistent with a giant planet or brown dwarf. The rounded bottom and depth in the primary eclipse of KID9772531 indicate an eclipsing object of radius $\sim$0.09$R_\sun \approx 0.9 R_J$, placing it on the intersection of giant planet, brown dwarf, and late M-dwarf star in the radius-mass relation of degenerate objects \citep{bt13, mon15}. KID11853130 has been included in a most recent compilation of planetary candidates (KOI-3263.01), albeit with a high false positive probability \citep{swi15}. 


One system, KID6023859, presents a clear and deep (8\%) primary but the secondary is absent. The fiducial period is 27 days but there is some evidence from RV data that the actual period could be twice this value (Jonathan Swift, private communications). If this is the case, the NSPS could be revised upwards by $\sim$0.8\%.   

High-resolution spectroscopy and RV monitoring with sensitivity to magnitudes $r\sim$16 should be able to elucidate the stellar nature and orbital elements of these companions.

\subsubsection{Higher-Order Multiples}

The existence and statistics of an outer hierarchical triple companion to short-period binaries have implications for their formation channels. We do not directly detect transiting higher-order companions among the EBs found in this study, hence do not attempt to place a constraint on its statistics. Dynamical stability and evolution arguments would suggest that, if triple companions existed, they are likely much farther out with the inner binary and not usually coplanar, hence are inaccessible via directly observing their transits in a survey of the current scope. Nonetheless, other methods may be used to infer the existence of further companions. For instance, from eclipse timing variations (ETVs), the presence of a third, non-transiting companion at $\sim3$ years has been inferred for one of the EBs (KID10979716) \citep{bor15}. KID3830820 is a blend of at least 3 stars, and these visual neighbours may be physically associated. The system of KID11853130 also harbours a visual companion at $0.8''$, identified through lucky imaging (PI Lillo-Box) and adaptive optics (PI Ciardi, see note 11 to \Table{EBlist}). This EB sample could benefit from more systematic high-resolution imaging to uncover triple companions at a few AUs.

~
\section{Conclusion}\label{sec:conclusion}

In the above exposition, we have described an exercise in mining data from the precision photometry monitoring survey \kep\ to conduct a study of M-dwarf multiplcity in close (but detached) orbits. The inference is rooted in the detection of EBs in this dataset, which has stared at $\sim3000$ field M-dwarfs as faint as {\it{r}}-magnitude 17 in the direction of Cygnus for nearly one year or more. The statistical method borrows directly from the calculation of exoplanet occurrence rates, with simplifications. Ours represents a very simple and straightforward way to infer the likelihood that an apparently single M-dwarf in the sky should harbour a companion within a 90-day orbital period. It uses a minimalist and transparent completeness correction, which suffices for its purpose given the nature of the technique and quality of the dataset. Moreover, this study involves real binary detections on the largest M-dwarf sample to date used for this type of statistical investigation.     

We find that, among the \kep\ M-dwarfs with no resolvable companion and whose photometry is consistent with that of a single star, there are on average $0.11 ^{+0.02} _{-0.04}$ (stellar) companions within a 90-day orbit. Assuming all apparently single M-dwarfs are either actually single or binary, this translates to a $11^{+2}_{-4}\%$ binarity fraction. This value is significantly higher than previous measurements of M-dwarf multiplicity from literature, though it may be partially inflated by potential survey biases (reflected in the lower bound error) as well as uncertainty regarding the nature of several objects, which could further cause 2-4\% overestimate. Two studies, FM92 and CBK12, both find a value under $4\%$ with $1-2\%$ uncertainty for a comparable binary separation regime using RV techniques. 

To reconcile the tension between the present work and the literature requires careful thought to the comparability of the studies, as discussed in \se{discuss}. RV followup to the binary detections presented in this paper would be critical to confirm or reject the stellar nature of companions and verify the interpretation of these systems. 

Future studies, both planned and in progress, will be key to refining this important measurement. One valuable venue could come from volume-limited surveys as that conducted by the RECONS team, which is a continual effort to exhaustively map the solar neighbourhood \citep[e.g.][]{hen06}. The statistics they will derive would form an interesting comparison to the several studies done on quasi-magnitude-limited surveys, and one should look forward to Winters et al. (in prep), which aims to quantify the M-dwarf multiplicity issue based on the RECONS data. In fact, according to a comprehensive private compilation of stars within 10pc of the Sun, there is already indication that the early M-dwarf population has a close binarity fraction greater than $5\%$ (Kevin Apps, private communications).      

\citet{des13} describes an ongoing effort to survey more than 1400 M-dwarf stars within the framework of the Sloan Digital Sky Survey, using the near-infrared APOGEE spectrograph. Preliminary data products are to be made public as part of the 12th SDSS data release. This project is intended to be useful for tackling questions of stellar multiplicity, among various other topics. The RV information collected over multiple epochs would enable sensitivity to a parameter space in orbital period that encapsulates that addressed in the present study.  

There is yet another potentially promising route that could help settle the issue of close stellar multiplicity, using photometric light curve data alone. It involves the use of distinct signatures of rotational modulation for the detection of multiple systems. An example of a possible implementation of this concept is described by \citet{rap14}.      

The present study also serves to preview the potence and urgency of ongoing and upcoming continuous photometry monitoring missions like the \kep\ 2-Wheeled Mission (K2) \citep{howell14} and the Transiting Exoplanet Survey Satellite (TESS) \citep{ric14}. \citet{kra15} recently studied in detail a mid-M+M eclipsing binary discovered with K2. Undoubtedly many will follow. Continuous microlensing surveys like the Wide Field Infrared Survey Telescope (WFIRST) \citep{wfirst} and the Korean Microlensing Telescope Network (KMTNet) \citep{atw12} will provide interesting opportunities to study multiple star populations in the direction of the galactic bulge. These programs have a critical role to play in the next decade of stellar astronomy.

\section{Acknowledgement}
Y.S. is grateful to her Research Exam Committee members Dave Charbonneau, Guillermo Torres, and Jonathan Irwin, for very helpful discussions and feedback at various stages of this project. She also thanks Andrew Vanderburg and Benjamin Montet for extensive suggestions on dealing with and interpreting \kep\ data. Cullen Blake, Brendan Bowler, Peter Plavchan, Jonathan Swift, Saul Rappaport, and Jason Wright have contributed important science insights and encouragement. Others who have played a valuable role in the conversation are Kevin Apps, Rebekah Dawson, Jason Dittmann, Courtney Dressing, Gaspard Duch{\^ e}ne, David Kipping, Dave Latham, Mercedes Lopez-Morales, Tsevi Mazeh, John Southworth, William Welsh, and Jennifer Winters. We thank the anonymous referee for their helpful feedback. J.A.J is supported by generous grants from the David and Lucile Packard Foundation and the Alfred P. Sloan Foundation. 

Some of the data presented in this paper were obtained from the Mikulski Archive for Space Telescopes (MAST). STScI is operated by the Association of Universities for Research in Astronomy, Inc., under NASA contract NAS5-26555. Support for MAST for non-HST data is provided by the NASA Office of Space Science via grant NNX09AF08G and by other grants and contracts. Specifically, this paper includes data collected by the \kep\ mission. Funding for the \kep\ mission is provided by the NASA Science Mission directorate. This paper has also made use of data from the \kep\ Community Follow-up Observing Program, maintained by Caltech's Infrared Processing \& Analysis Center (IPAC).

\vspace{+10pt}

\end{document}